\documentclass[twocolumn,10pt]{extarticle}
\usepackage{achemso}
\usepackage{jpc}
\usepackage{graphicx} 
\usepackage{amssymb}
\usepackage{amsmath}
\usepackage{color}
\usepackage{natbib}
\usepackage{graphicx} 
\usepackage{amssymb}
\usepackage{amsmath}
\usepackage{color}
\usepackage{comment}
\usepackage[normalem]{ulem}
\usepackage{wasysym}
\usepackage{bm}
\usepackage[pdftex,colorlinks,citecolor=blue,urlcolor=blue]{hyperref}
\usepackage{changepage}
\usepackage{multirow}
\usepackage{threeparttable}
\title{An accurate alternative to hybrid functionals for germanium: DFT+\texorpdfstring{$\alpha$}{a}}

\begin{document}
\author{Abdulgaffar Abdurrazaq,$^{a,b}$ Ruggero Lot,$^{a,c}$ Antoine Jay,$^{a}$
Gabriela Herrero-Saboya,$^{d,*}$\\
Nicolas Richard,$^e$ Layla Martin-Samos,$^d$ Anne H\'emeryck,$^a$ Stefano de Gironcoli$^{b,d}$\\
\small $^a$LAAS-CNRS, Université de Toulouse, CNRS, F-31400 Toulouse, France\\
\small $^b$SISSA–Scuola Internazionale Superiore di Studi Avanzati,  I-34136 Trieste, Italy\\
\small $^c$AREA Science Park,  I-34149 Trieste, Italy\\
\small $^d$CNR-Istituto Officina dei Materiali (IOM), c/o SISSA, I-34136 Trieste, Italy \\
\small $^e$CEA, DAM, DIF, F-91297 Arpajon, France\\
} \date{}

\twocolumn[
\maketitle
\begin{abstract} 
The accuracy of bulk property predictions in density functional theory (DFT) calculations depends on the choice of exchange-correlation functional. While the Perdew–Burke–Ernzerhof (PBE) functional systematically overestimates lattice parameters and strongly underestimates electronic band gaps, hybrid functionals such as Heyd–Scuseria–Ernzerhof (HSE) offer better overall agreement across a broad range of materials. Using germanium as a critical test case, we challenge the ability of both functionals to capture semiconductor properties. Although HSE improves PBE’s gap error, it fails to reproduce germanium’s correct $\Gamma$-L indirect and  $\Gamma$-$\Gamma$  band gaps simultaneously. 
Noting that the PBE underestimated energy separation between the  4p valence-band maximum and 4s conduction-band minimum causes unphysical $sp$ mixing, we propose DFT$+\alpha$, a semi-empirical correction scheme applied selectively to 
4s-like orbitals. For germanium, DFT$+\alpha$ restores the proper ordering and orbital character of the band edges and  yields accurate lattice constant, bulk modulus, elastic constants and phonon frequencies at a fraction of hybrid-functional computational cost.
\end{abstract}
\vspace*{2em}] \thanks{\noindent
  \footnotesize
   *Corresponding Author: Gabriela Herrero-Saboya \\ 
  E-mail: herrero@iom.cnr.it 
}
\section{Introduction}
The accuracy of bulk property predictions in density functional theory (DFT) calculations is determined by the choice of exchange-correlation functional. Extensive benchmarking efforts have evaluated the performance of various functionals across a wide range of materials\cite{Borlido:2019}.
These studies highlight the limitations of the Perdew-Burke-Ernzerhof (PBE) functional~\cite{Perdew:1996}, which systematically overestimates  lattice parameters and underestimates electrical band gaps.
Hybrid functionals, such as Heyd-Scuseria-Ernzerhof (HSE), modify standard exchange-correlation approximations by incorporating a fraction of Hartree-Fock exchange along with a screening parameter that controls the range of exchange interactions~\cite{Heyd:2003,Heyd:2006}. 
With adjusted parameters, HSE and other hybrids offer a practical trade-off between computational efficiency and overall accuracy across a broad range of materials.
However, when focusing on a specific system, it may fail to deliver broad predictive accuracy.\\

For semiconductors, HSE has become the reference computational approach.
Germanium, with an indirect band gap ($\Gamma$-L) of 0.74 eV and a  $\Gamma$-$\Gamma$  band gap of 0.90 eV~\cite{Madelung:2004}, provides a critical test for HSE's consistency in accurately describing bulk properties.
Indeed, PBE severely underestimates both band gaps, often predicting a metallic system or gaps as small as a few meV.
While HSE systematically corrects the band gaps to the right order of magnitude, reported values are significantly dispersed, with discrepancies in the absolute and relative energies of both band gaps~\cite{Martin:2006,Hummer:2009,Deak:2010,Weber:2013,Pasquarello:2016}.
In most cases, the  $\Gamma$-L band gap is smaller than the  $\Gamma$-$\Gamma$ one, but both values frequently remain close to degeneracy. 
When compared to the experimental band gaps at low temperatures (T = 1.5 K)~\cite{Madelung:2004},
only a few calculations reproduce the $\Gamma$-L band gap, while underestimating the $\Gamma$-$\Gamma$ band gap by at least 10\%~\cite{Martin:2006,Hummer:2009,Deak:2010,Weber:2013,Pasquarello:2016}.\\

HSE band gap predictions remain scattered even when using the same code and pseudopotential implementation, identical exchange mixing and screening parameters, and no spin-orbit coupling~\cite{Hummer:2009, Deak:2010, Weber:2013}.
Two of these studies attempt to refine the indirect band gap further, one by incorporating spin- orbit coupling corrections~\cite{Hummer:2009},
the other by adjusting the exchange mixing parameter~\cite{Weber:2013}.
Notably, in the work of Peralta \textit{et al.}~\cite{Martin:2006}, spin-orbit coupling corrections alone failed to reproduce the correct relative energy between the band gaps, requiring an adjustment of the lattice parameter to its experimental value.\\

In this work, we challenge the ability of PBE and HSE functionals to describe semiconductor properties. We show that even though  HSE mitigates PBE’s severe underestimation of germanium's $\Gamma$-L and $\Gamma$-$\Gamma$ band gaps, it fails to accurately reproduce both band gaps.
 Seeking an alternative approach, we reconsider the PBE band structure, where the energy difference between the valence band maximum with 4p character and the conduction band minimum with 4s character is significantly underestimated, resulting in a sp nonphysical mixing. Exploiting this specific aspect, we introduce DFT$+\alpha$, a semi-empirical correction scheme that selectively shifts the energy of 4s-like orbitals, recovering both the correct band edges alignment and their corresponding orbital character.
 This refinement of germanium’s electronic structure enables an accurate reproduction of the lattice parameter, bulk modulus $B_0$, elastic constants ($C_{11}$, $C_{12}$ and $C_{44}$) and  phonon frequencies, offering a robust, computationally inexpensive alternative to standard hybrid functionals. 

\section{Computational details}
\label{sec:comp}
All calculations of germanium in its diamond lattice, with \textit{Fd$\bar{3}$m} space group symmetry, were performed using the Quantum ESPRESSO package for electronic structure calculations~\cite{Giannozzi:2017}.
The germanium unit cell is described using a PAW non-relativistic pseudopotential from the PseudoDojo platform~\cite{Jollet:2014, Ge-PP} that includes 3d electrons.
A plane-wave basis set is used with a kinetic energy cutoff of 80 Ry and a density cutoff of 320 Ry.
The Brillouin zone (BZ) is sampled with a 12$^3$ Monkhorst-Pack grid.
The optimized lattice parameters were determined by finding the minimum of the energy-volume curve.
A smearing of 0.001 Ry is imposed to ensure convergence of the metallic states when using the PBE functional. In addition,  HSE calculations were done with a 4$^3$ q-grid and a Fock operator plane-wave cutoff equal 320 Ry. 
Elastic constants have been calculated using the termo\_pw package included in Quantum ESPRESSO. The phonon dispersion curves are calculated using PHONOPY~\cite{Togo:2023} and a 5$^3$ supercell with one displaced atom.

\section{Results}
We perform DFT calculations for germanium using the HSE functional as implemented in the Quantum ESPRESSO package~\cite{Giannozzi:2017},
with default values of 0.106 Bohr$^{-1}$ and 0.25 for the screening and exchange mixing parameters, respectively.
Shown in Figure~\ref{fig:HSE} are the $\Gamma$-$\Gamma$ and $\Gamma$-L band gaps, the lattice constant and the elastic constants $B_0$, $C_{11}$, $C_{12}$ and $C_{44}$ for exchange mixing parameters ranging from 0.21 to 0.35.
For each exchange coefficient, both the electronic and elastic properties are determined at the optimized lattice constant.
For the standard exchange parameter, 0.25, we obtain a lattice constant of 5.71~\AA~ (Figure~\ref{fig:HSE}b) and $\Gamma$-$\Gamma$ and $\Gamma$-L band gaps of 0.775 and 0.823~eV, respectively (Figure~\ref{fig:HSE}a).
The lattice constant differs by $\sim$1.02~\% from the experimental value of 5.652~\AA~ measured at low temperatures (T=8~K)~\cite{Hu:2003}.
The computed band gaps also deviate from experimental data, both in their absolute and relative values~\cite{Madelung:2004}.
When the exchange parameter is increased beyond 0.28, the magnitudes of the $\Gamma$-$\Gamma$ and $\Gamma$-L band gaps become inverted, but their absolute values still show discrepancies of 34~\% and 17~\%, respectively, for an exchange coefficient of 0.3.
Regarding the elastic properties (Figure~\ref{fig:HSE}c),  the bulk modulus and the $C_{44}$ constant remain relatively close to the  experimental references of 75 GPa and 68 GPa~\cite{Madelung:2004}, respectively, for all exchange parameters. However, at the standard exchange parameter,  $C_{11}$ and $C_{12}$ differ by about 15\% and 35\%, respectively, from their experimental values of 124 GPa and 41 GPa~\cite{Madelung:2004}.\\

\begin{figure}[thb!]
\includegraphics[width=0.49\textwidth]{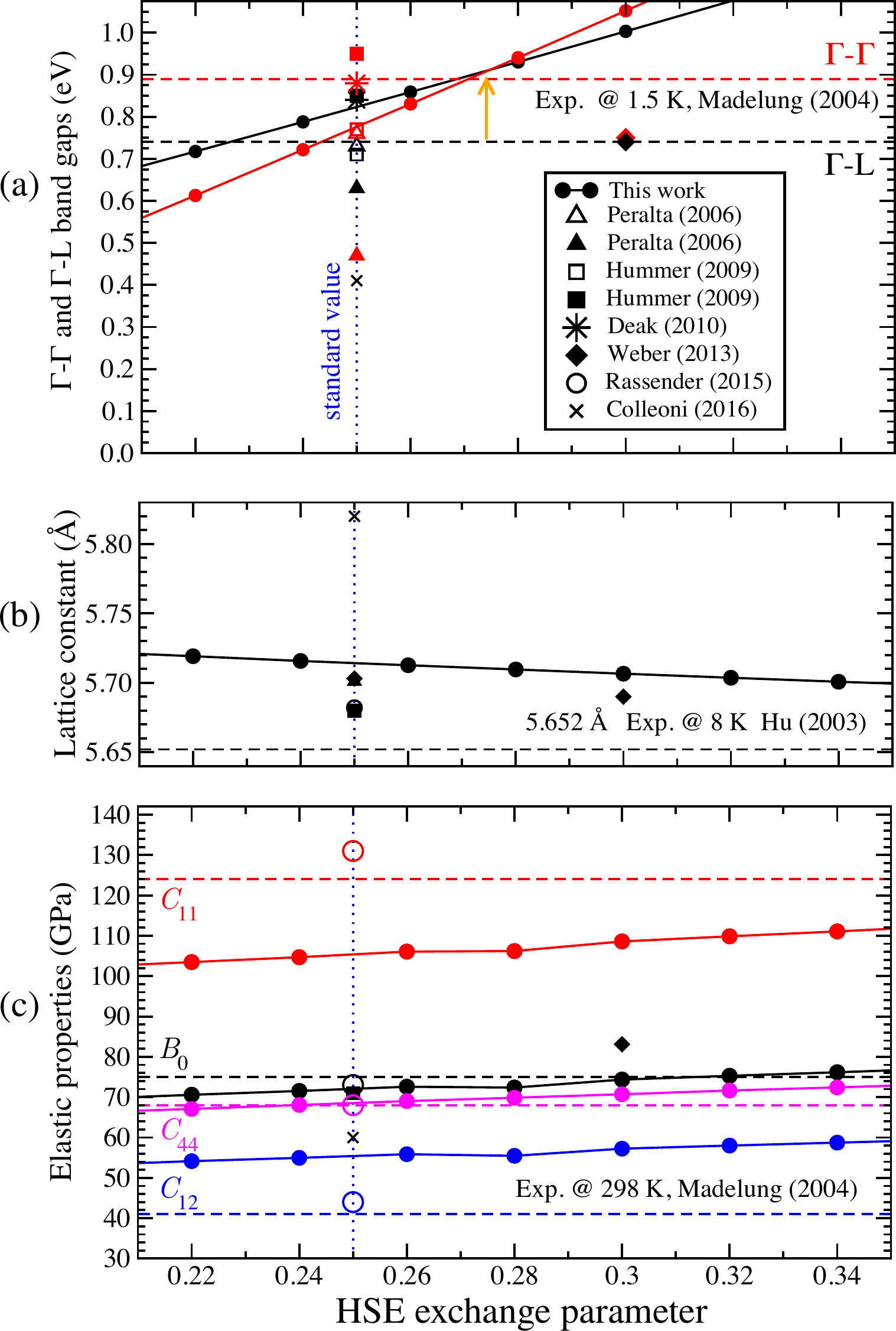}
\caption{\label{fig:HSE}
DFT calculations of germanium's properties using the HSE functional, displayed as a function of the exchange mixing parameter.
(a) The $\Gamma$-$\Gamma$ and $\Gamma$-$\text{L}$ band gaps calculated at the optimized lattice parameter. 
Experimental values are taken from ref~\citep{Madelung:2004} and previous results from refs.~\citep{Martin:2006,Hummer:2009,Deak:2010,Weber:2013,Pasquarello:2016}.
(b) The optimized lattice parameter.
Exp. from ref.~\citep{Hu:2003} and previous results from refs.~\citep{Martin:2006,Hummer:2009,Weber:2013,Rasander:2015,Pasquarello:2016}.
(c) The elastic constants $B_0$, $C_{11}$, $C_{12}$ and $C_{44}$.
Experimental constants are taken from ref.~\citep{Madelung:2004} and previous results from refs.~\citep{Hummer:2009,Weber:2013,Rasander:2015,Pasquarello:2016}.}
\end{figure}

Overall, our HSE results deviate from the experimental data~\cite{Madelung:2004,Hu:2003}, a limitation shared by all previous HSE calculations~\cite{Martin:2006,Hummer:2009,Deak:2010,Weber:2013,Rasander:2015,Pasquarello:2016}, which exhibit scattered reproducibility and uneven accuracy (Figure~\ref{fig:HSE}).\\

Seeking an alternative approach, 
we reconsider the PBE's band structure. 
Unlike  diamond and silicon, germanium's sp$^3$ hybridization results in a valence band maximum with a 4p character
at the center of the Brillouin zone, while the conduction
band minimum exhibits a 4s character.
In the PBE approximation, the energy difference between these bands is significantly underestimated, resulting on an sp nonphysical mixing. 
Exploiting this specific aspect, we introduce DFT+$\alpha$, a semi-empirical correction scheme that selectively shifts the energy of 4s-like orbitals, recovering both the correct band edges alignment and their corresponding orbital character. \\

Under the DFT+$\alpha$ scheme, the eigenvalues of the Kohn-Sham Hamiltonian have been modified as:
\begin{equation}
\label{eq:alpha}
E \left[ n(\mathbf{r}) \right] = E_\text{KS} \left[ n(\mathbf{r}) \right] + \alpha \sum_{I, \sigma} \text{Tr} 
\left[ n^{I \sigma}(\mathbf{r}) \right],
\end{equation}
where $E_\text{KS} \left[ n(\mathbf{r}) \right]$ is the standard Kohn-Sham energy for the electronic density $n(\mathbf{r})$ 
and $n^{I \sigma}(\mathbf{r})$ is the occupation matrix of the selected $4s$-like states, 
located in atom $I$ and spin $\sigma$.
The $\alpha$ coefficient modulates the energy increase of the targeted Kohn-Sham states.\\

\begin{figure}[thb!]
\includegraphics[width=0.49\textwidth]{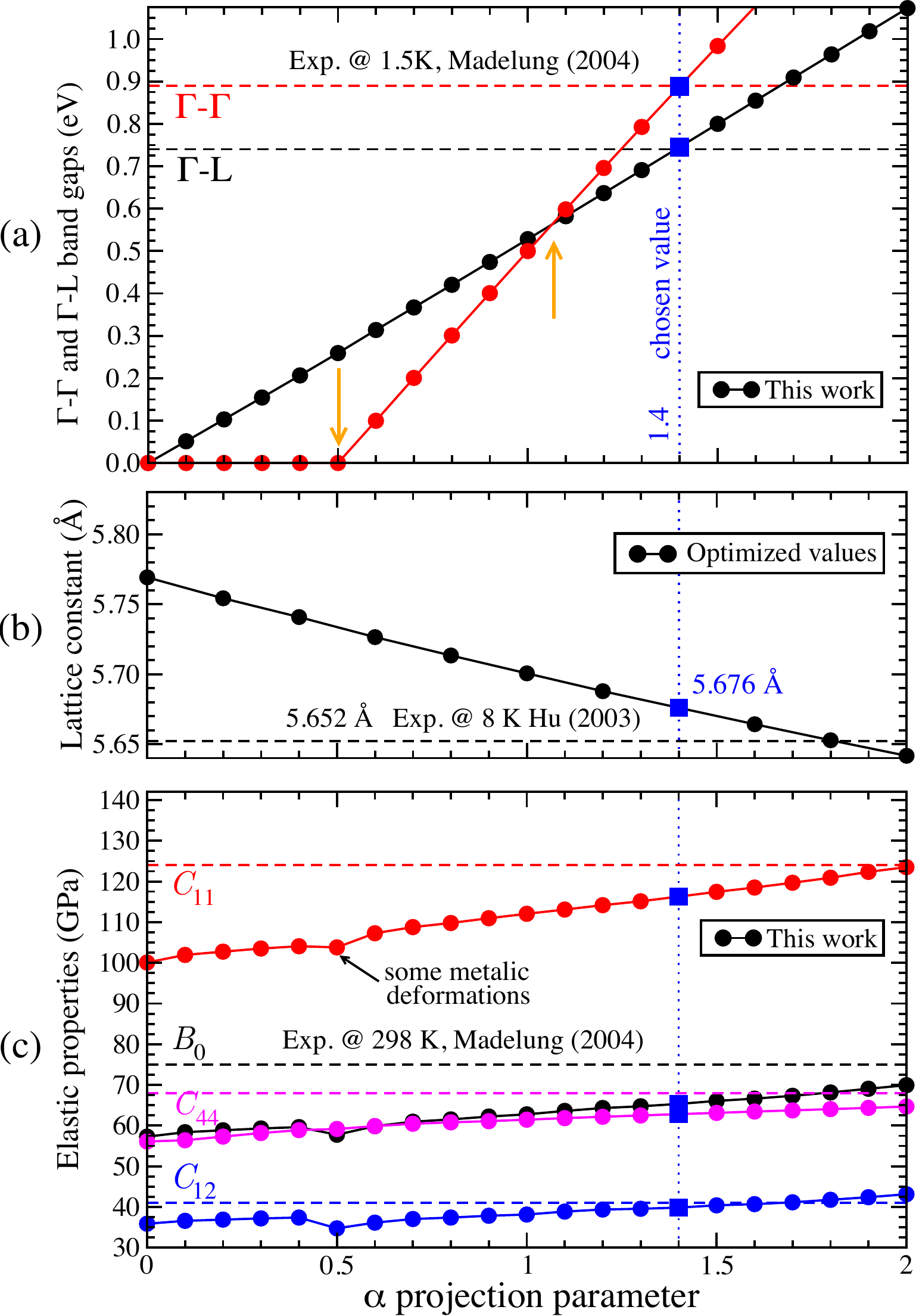}
\caption{\label{fig:DFTa}
DFT+$\alpha$ calculation of germanium's properties using the PBE functional displayed as a function of the $\alpha$ parameter.
(a) The $\Gamma$-$\Gamma$ and $\Gamma$-$\text{L}$ band gaps at the optimized lattice parameter for each $\alpha$ value.
Yellow arrows mark the transition from a closed band gap to an open one, as well as the inversion of the $\Gamma$-$\Gamma$ and $\Gamma$-L band gaps.
Experimental values are taken from ref.~\citep{Madelung:2004}.
(b) The optimized lattice parameter.
The experimental lattice parameter is reported in ref.~\citep{Hu:2003}.
(c) The elastic constants, $B_0$, $C_{11}$, $C_{12}$ and $C_{44}$ at the optimized lattice parameter for each $\alpha$ value.
Experimental constants are collected from ref.~\citep{Madelung:2004}.
}
\end{figure}

Within the DFT+$\alpha$ method, defining the occupation matrix, $n^{I \sigma}(\mathbf{r})$, requires specifying the occupation of atomic orbitals in a multi-atomic system.
Given the inherent arbitrariness of this task, possible approaches include defining these occupations based on  projections on atomic orbitals or Wannier functions. 
The occupation matrix can be written in the following generic form,
\begin{equation}
\label{eq:occ_matrix}
n^{I\sigma} = \sum_{\mathbf{k} v} f^{\sigma}_{\mathbf{k} v} \left< \psi^{\sigma}_{\mathbf{k} v} | \hat{P}^{I} | \psi^{\sigma}_{\mathbf{k}v} \right>,
\end{equation}
where $\psi^{\sigma}_{\mathbf{k}v}$ is the KS wave function of the state $\mathbf{k}v$ and spin $\sigma$ 
and $f^{\sigma}_{\mathbf{k} v}$ is the corresponding occupation.
$\hat{P}^{I}$ is the projector operator on the wave function $\psi^{I}$ of atom $I$,
defined as $\hat{P}^{I} = \sum_{I} |\psi^{I}\left> \right< \psi^{I}|$.
We define $\psi^{I}$  as,
\begin{equation}
\label{equ:projbase} 
 \psi^{I} (r) = \text{C} [(1+2r^2 + \frac{1}{2}(2r^2)^2 + \frac{1}{6}(2r^2)^3]e^{-2r^2} \phi_s (r)
\end{equation}
where $\phi_s (r)$ is the  4s atomic orbital of the germanium pseudopotential and C is the normalization factor.
This normalized function primarily corresponds to the 4s orbital of the pseudopotential, 
but its long-range tail has been suppressed with the exponential term 
to minimize the overlap between wave functions of neighboring atoms. 
This modified pseudopotential is used in the localized manifold projector,  introduced to compute the effective interaction parameters in the $\mathrm{LDA}+\mathrm{U}$ method~\cite{Cococcioni:2005}, as implemented in the Quantum ESPRESSO (QE) package~\cite{Giannozzi:2017}. Although DFT+$\alpha$ is proposed for the critical case of germanium, this scheme can also be applied to related semiconductors that exhibit sp mixing.\\

As the $\alpha$ parameter modifies the alignment between 4s and 4p orbitals, it must be tuned to reproduce the band gaps. 
We demonstrate that by adjusting only this single 
$\alpha$ parameter, one can achieve consistent accuracy across all bulk properties.
As shown in Figure~\ref{fig:DFTa}a, both the $\Gamma$-$\Gamma$ and $\Gamma$-L band gaps are estimated for different $\alpha$ values, ranging from 0 (corresponding to a standard PBE calculation) to 2.
The optimized lattice parameter was used to calculate the pair of band gaps for each $\alpha$ value.
As illustrated by the yellow arrows, the $\Gamma$-$\Gamma$ band gap opens for $\alpha$ greater than 0.5 and inverts in magnitude with the $\Gamma$-L gap for values above 1.07. 
For $\alpha=1.4$ (blue dotted line in Figure~\ref{fig:DFTa}a),
both band gaps are within less than 2\% of the experimental values~\cite{Madelung:2004}.
For this $\alpha$, the optimized lattice constant is 5.676~\AA, which is within 0.4\% of the experimental reference (Figure~\ref{fig:DFTa}b).
For comparison, a standard PBE calculation finds as optimal value $a$=5.769~\AA, a lattice constant with more than 2\% of relative error.
For the electronic gaps and the lattice parameter,  increasing the $\alpha$ value forces the opening of the $\Gamma$-$\Gamma$ gap and consequently decreases the lattice parameter.\\

\begin{figure}[thb!]
\includegraphics[width=0.49\textwidth]{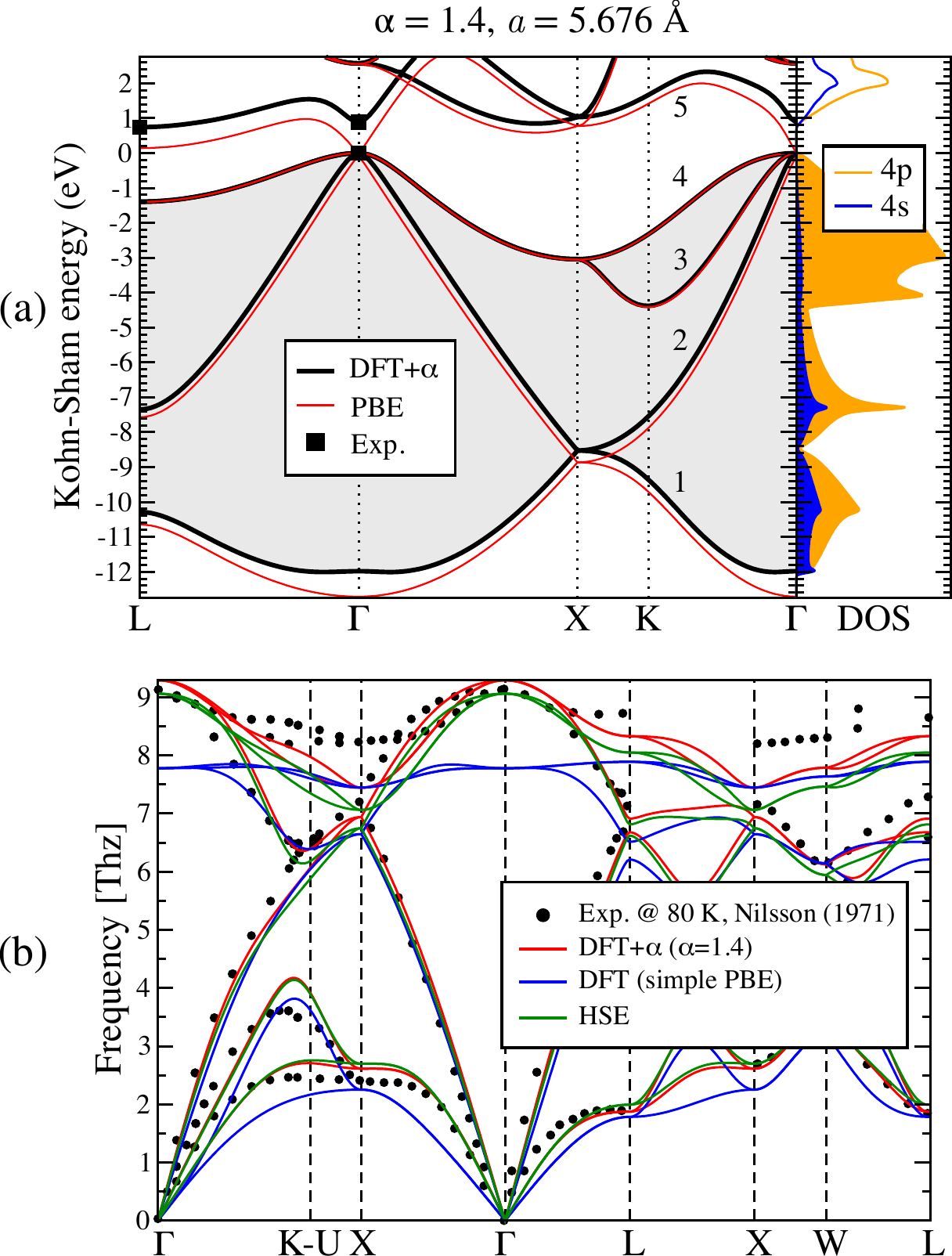}
\caption{\label{fig:DFTa_bands}
DFT+$\alpha$ calculation of germanium's properties for the chosen value $\alpha = 1.4$ at its optimized lattice constant, $a = 5.676$ \text{\AA}.
(a) Electronic band structure within the DFT+$\alpha$ approach is shown in black.
For comparison, the standard DFT calculation using the PBE functional has been performed at the same lattice parameter and is shown in red.
Gray-filled areas correspond to occupied states or bands 1-4.
Experimental values for the $\Gamma$-$\Gamma$ and $\Gamma$-L band gaps at cryogenic temperatures are taken from ref.~\citep{Madelung:2004}.
The projected electronic density of states (DOS) onto the 4s and 4p atomic orbitals for the DFT+$\alpha$ calculation is also included. 
(b) Phonon dispersion curves of Ge, computed using standard DFT (blue), DFT+$\alpha$ (red), and HSE (black).
For each calculation, the corresponding optimized lattice constant is used. 
Experimental values are taken from ref.~\citep{Nilson:1971}.}
\end{figure}

For a more in-depth understanding of the effect of $\alpha$ on the KS eigenvalues, we plot the germanium band structure using standard PBE and PBE+$\alpha$ calculations with $\alpha$=1.4 (Figure~\ref{fig:DFTa_bands}a).
For each calculation, the optimized lattice parameter of PBE+$\alpha$ is used.
As expected, the first and fifth bands are the most affected by the DFT+$\alpha$ formulation, as they have the \textit{highest 4s character}. 
However, their shift relative to the standard KS eigenvalues varies along the BZ, depending on the 
degree of hybridization between the 4s and 4p atomic orbitals. 
At $\Gamma$, the first and fifth bands are purely 4s, leading to the largest deviation between PBE and PBE+$\alpha$, effectively opening the $\Gamma$-$\Gamma$ band gap to the experimental value~\cite{Madelung:2004}.
\\

Though the DFT+$\alpha$ method was developed to improve germanium's electronic properties, 
it also proves to be a robust method for describing the ground state properties of the bulk, such as the lattice constant (Figure~\ref{fig:DFTa}c).
Moreover, as shown in Figure~\ref{fig:DFTa}c, the elastic constants $B_0$, $C_{11}$, $C_{12}$ and $C_{44}$ are in close proximity to experimental references, with a relative error lower than 6\%. The discontinuity in the $B_0$($\alpha$) and $C_{ij}$ ($\alpha$) curves slightly above $\alpha$=0.45 is due to the cell deformations used to calculate the elastic constants, which open or close the $\Gamma$-$\Gamma$ band gap, leading to irregular fits of the equations of state.
For comparison, standard PBE values exhibit relative errors in the range of 19-24\%. \\

Regarding the vibrational properties of germanium, the PBE+$\alpha$ approach also addresses the limitations of the PBE description of phonon frequencies.
In Figure~\ref{fig:DFTa_bands}b, we plot the phonon dispersion curves calculated using PBE, PBE+$\alpha$ ($\alpha=1.4$) and HSE calculations against neutron diffraction data~\cite{Nilson:1971}.
For each calculation, the optimized lattice parameter is used.
For the acoustic branches, PBE+$\alpha$ aligns well with the HSE calculations, significantly improving the description of the branches' slopes near $\Gamma$, \textit{i.e.}, the sound velocities.
For the optical modes, both PBE+$\alpha$ and HSE increase the underestimated frequencies, aligning well with experimental data near $\Gamma$.
However, at the edges of the BZ, both approaches tend to underestimate the optical branches. In the case of PBE+$\alpha$, this misalignment may result from the lack of piecewise linearity in the approach and/or from an inaccurate description of long-range interactions. We note that, although our scheme improves the description of bulk properties by manually correcting the sp mixing, it does not remedy the intrinsic limitations of exchange–correlation functionals.

\section{Discussion}
In summary, we challenge the conventional understanding of PBE and HSE functionals in accurately describing semiconductor properties.
For germanium, we have demonstrated that while HSE corrects PBE’s severe underestimation of the $\Gamma$-L and $\Gamma$-$\Gamma$ band gaps, it fails to reproduce both experimental values simultaneously.
As an alternative, we propose the DFT$+\alpha$ approach, a tailored correction that adjusts the erroneous energy difference between germanium's 4s and 4p bands to effectively open the band gap.
We demonstrate that by adjusting only this energy difference, one can achieve consistent accuracy across all bulk properties.\\

Beyond the description of physical properties, DFT+$\alpha$ provides a computational advantage over hybrid calculations, as it scales comparably to a standard self-consistent field (SCF) calculation. For germanium, this approach opens the band gap, enabling the accurate and yet computationally efficient exploration of its ground state-properties. \\

While DFT+$\alpha$ is a semi-empirical approach, it addresses a major drawback of the most commonly used exchange-correlation functionals: the sp mixing.  For germanium, this correction is critical, transforming its description from metallic to semiconducting.  Since sp mixing in diamond and zinc-blende lattices is a recurring limitation,
it also provides improvements for related semiconductors. \\


\section*{Data availability}
The data that support the findings of this study are available upon reasonable request.

\section*{Code availability}
The Quantum ESPRESSO package and the PHONOPY code are open-source software that can be obtained from https://www.quantum-espresso.org/ and https://phonopy.github.io/phonopy/ respectively.

\section*{Acknowledgments}
This work was carried out using HPC resources from CALMIP-Grant P1555. This work was supported by the French GeSPAD Project of the Agence Nationale de la Recherche (ANR-20-CE24-0004). This work was also supported by the European Commission through the MAX Centre of Excellence for supercomputing applications (grant numbers 10109337 and 824143) and by
the Italian MUR, through the Italian National Centre from HPC, Big Data, and Quantum Computing (grant number CN00000013).

The authors gratefully acknowledge the technical support of Pietro Delugas.

\bibliography{mybiblio}

\end{document}